\documentclass[journal]{IEEEtran}
\usepackage{blindtext}
\usepackage{graphicx}
\usepackage{subcaption}
\ifCLASSINFOpdf
\else
\fi

\usepackage{paralist}

\begin{document}
%
\title{A Framework for Semantic In-network Caching\\ and Prefetching  in 5G Mobile Networks}
%
%
%

\author{Can~Mehteroglu,~Department of Computer Engineering,~METU,
        Yunus~Durmus,~Accenture,~The Netherlands
        and~Ertan~Onur,~Department of Computer Engineering,~METU
\thanks{C. Mehteroglu and E. Onur are with the WINS Lab at the Department
of  Computer Engineering, Middle East Technical University (METU), 06800 Ankara, Turkey e-mail: eronur@metu.edu.tr.}
\thanks{Y. Durmus is with Accenture, the Netherlands.}
\thanks{This work is partially presented in IEEE CCNC 2017 and is the output of the MS thesis of Can Mehteroglu at METU.}}

\maketitle

\begin{abstract}
Recent popularity of mobile devices increased the demand for mobile network services and applications that require minimal delay. 5G mobile networks are expected to provide much lesser delay than the present mobile networks. One of the conventional ways for decreasing latency is caching content closer to end users. However, currently deployed methods are not effective enough. In this paper, we propose a new astute in-network caching framework that is able to smartly predict subsequent user requests and prefetch necessary contents to remarkably decrease the end-to-end latency in 5G mobile networks. We employ semantic inference by  edge computing, deduce what the end-user may request in the sequel and prefetch the content. We validate the proposed technique by emulations, compare it with the state of the art and present impressive gains.
\end{abstract}

\begin{IEEEkeywords}
Semantic caching, prefetching, 5G, in-network caching, content-aware networking
\end{IEEEkeywords}

%
\IEEEpeerreviewmaketitle

\section{Introduction}

Not only does the number of  mobile devices  increase  but also  capabilities of  them advance rapidly. They are no longer limited  to voice communication and text messaging. Internet multimedia browsing, location tracking, or collecting sensory information  are considered as typical features. This trend imposes  huge data rate and minimal latency requirements on next generation  communications systems. 5G is foreseen to provide a ``zero latency gigabit experience.''  Reduction of latency is considered as equally important as increased  rates. Zero latency does not  mean no delay but it implies delay should be lower than 1 millisecond \cite{nokia5GUseCases}.

Since an important percentage of mobile traffic is a consequence of duplicated downloads of the same content, efficacious caching strategies to prevent redundant transmissions are to be developed \cite{cacheInTheAir}. According to Liu et al., 10 to 70\%  redundant traffic is produced when accessing  streaming services from iOS devices \cite{Liu2014}. When  data are served from  caches located closer to  end users, latency may become smaller. If  routers and other elements in a (mobile) network were capable of inferring what  subsequent requests of users will be, they may prefetch and cache the responses before the actual request originates. Consequently, the time required for serving the data  when the actual request originates can be reduced and  redundant traffic can be eliminated.  Towards this aim, we propose a smart prefetching and  in-network caching framework using semantic inference technologies. It allows network elements such as  routers to understand  meaning of packets and take action based on the meaning.  Our main objective is not decreasing the latency below one millisecond but facilitating the network to achieve this goal by providing an effective content prefetching and caching framework. 

\begin{figure}[t!]
\centering
\includegraphics[width=\columnwidth]{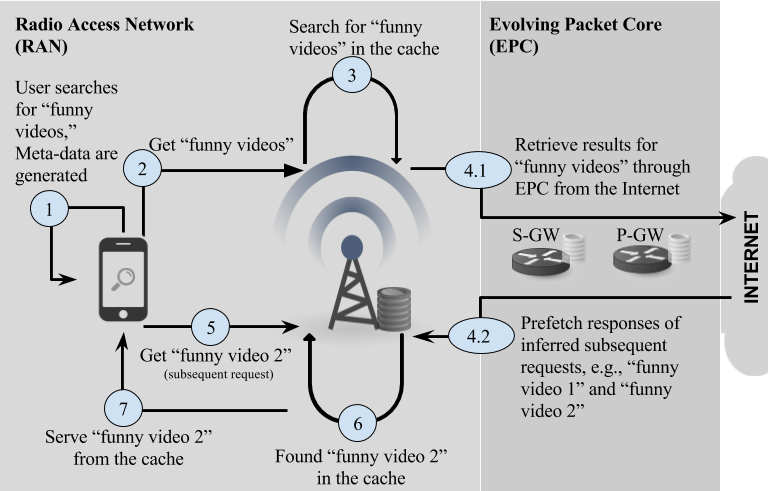}
\newline
\caption{An example scenario depicting the semantic caching and prefetching framework.}
\label{figure_3}
\end{figure}

To facilitate routers to reason on the requests and to make associations among requests in  a collaborative fashion, the meaning of each request must be known and understood by the routers that have some computation power such as  cloudlets as shown in Figure~\ref{figure_3}. The meta-data representing the meaning of the content have to be transmitted to the network elements. Most of the existent caching strategies employ application layer protocols on top of end-to-end transport layer protocols. Application or transport layer packets  cannot be processed by routers especially when the payloads are encrypted. The meta-data have to be conveyed to the routers in the Internet Protocol (IP) datagrams. We propose transportation of meta-data in the extension headers of IPv6 in this framework.  When the meta-data arrive at the routers, they will be able to make inferences using ontologies. We do not concentrate on the problem of generating meaning of the data but propose a framework for transportation of the semantic meta-data between hosts and the routers as well as using it in the routers.

An example scenario that illustrates the framework is shown in Figure \ref{figure_3}. A user queries ``Funny Videos'' using a search engine and  an HTTP request is generated and the meta-data describing the request are inserted in the hop-by-hop options header of the IPv6 datagram (step 1). That request is transferred to the eNodeB (step 2). In the cloudlet at the base station (eNodeB), the meta-data are extracted from the options (extension) headers and searched in the cache (step 3). We assume  the request in not found in the cache. Then, the HTTP request is forwarded to the Internet and the response is retrieved in step 4.1. While the HTTP result is fetched from the Internet and sent to the user, similar contents to ``Funny Videos" are found making inferences using the meta-data at the  cloudlet. Related contents that may be the subsequent request of the user could be one of the links that will appear on the first page of ``Funny Videos" query result depending on the used semantic inference algorithm described in Section \ref{atTheMobileEdgeCloud}. We assume here  the semantic inference algorithm has chosen videos appeared on the first page. Then, these contents are fetched from the Internet and stored in the cache with their meta-data. Prefetching is marked as step 4.2 that runs simultaneously with step 4.1. The subsequent request of the user shown in step 5 reaches to the cloudlet. A content with this meta-data will be searched and found in the cache, which is step 6, because whole videos on the first page of ``Funny Videos" result has already been prefetched and cached. The content is served to the user directly from the cache without traveling through the Internet and mobile core. This will significantly reduce the end-to-end latency (step 7). The bandwidth consumed for  meta-data transportation in step 2 and  for prefetching in step 4.2 comprise the cost portion of the trade-off for reduced latency in step 7. However, if the user makes a request for any other video on the first page, it will be served from the cache without causing any additional costs. Moreover, if another user in the same cell wants to watch one of the videos that are on the first page of ``Funny Videos" request, that user will retrieve it from the cache without any additional cost as well that will reduce the overall delay in the system and redundant traffic.

The semantic prefetching and caching framework that we propose in this paper
\begin{inparaenum}[ a)\upshape]
\item  is content-based, it knows about the \textit{meaning} of content by using semantic technologies,
\item  makes inferences on each user request and predicts future user requests,
\item  is an in-network caching strategy that serves content from caches even at the first time it is demanded
\item  is proactive, it prefetches contents before they are requested, and
\item  facilitates collaborative filtering and caching.
\end{inparaenum}
Some of these properties are also valid for other methods in the literature. \textit{Making inferences and prefetching contents based on those inferences} on each user request is the main contribution of this framework. Since the meta-data are carried in the extension headers of IP datagrams, routers are able to process the meaning of payloads and in-network caching becomes probable. Not only is the meaning of payload  carried, but also  user, device and context information can be included in the meta-data that may enable collaboration and innovative applications.  

After presenting traditional caching strategies briefly in the next section, we present the proposed framework in Section~\ref{sec:semanticcaching}. The validation results and discussion are presented in Section~\ref{sec:validationAndResults} and then we conclude the paper.

\section{Traditional Caching Strategies}
\label{sec:traditionalcaching}

Web caching takes the advantage of the fact that mobile traffic employs HTTP for 82\% of the time \cite{toCacheOrNotToCache}. Web caching identifies content using their uniform resource locators (URL). It cannot avoid duplicate transmission of the same content when they have different URLs. Temporary content (i.e., one time used content) cannot be cached  and content updates are not observed with standard web caching methods \cite{comparisonOfCaching}.

Byte caching \cite{byteCaching} transmits only the fingerprint of the contents that are being retransmitted. Since fingerprints are very small in size, a prominent amount of bandwidth is saved and  latency is decreased. Byte caching caches the content after dividing it into smaller chunks and each chunk has its own fingerprint. This means that, byte caching is able to detect and benefit from partial overlap in similar contents. However, it is not able to predict any user event and cannot serve any content from the caches when it is requested for the first time.

A proactive caching method called Proactive User Preference Profile (P-UPP) It prefetches videos that are most likely to be requested by the users in a cell by calculating  preference profiles of users \cite{videoCaching}. Caches are located at eNodeBs.  P-UPP can serve content even at the first request.  Since the caches are located at the radio access network (RAN),  latency will be considerably small. In P-UPP, upon user arrivals or departures in a cell, video request probabilities are recalculated. If expected improvement of cache hit ratio is greater than a threshold, necessary video contents are pre-loaded to the caches. There is no prefetching at user events other than location changes. After a user makes a request, this method does not prefetch any content using the information related to that request. It does not benefit from the fact that a request coming from a user may signal other requests and  it does not consider cooperation opportunities such as collaborative filtering.

An effective approach for predicting future requests is Markov-based prefetching \cite{aSurveyOfWebCachingAndPrefetching}. It predicts subsequent requests of a user by matching the user's current surfing path with  historical surfing paths. Surfing paths observed in the past may indicate future surfing paths. Therefore, this method cannot predict a future request that has never been made in the past. Markov-based prefetching is not able to serve any content from the cache at the first time it is requested \cite{aSurveyOfWebCachingAndPrefetching,webLatencyReductionWithPrefetching}. 

Similar to Markov-based prefetching, data mining-based prefetching employs history information \cite{aSurveyOfWebCachingAndPrefetching}. It is classified into rule- or clustering-based  approaches. Approaches based on association rules  may cause inconsistent predictions with large datasets \cite{anIntegratedModel}. An example for clustering-based prefetching  is the rank-based prefetching method presented in \cite{pageRankBased}. This method ranks the web pages linked to the requested page and prefetches the pages with the highest ranks. The main disadvantage of the method is that it provides low hit rates with non-clustered requests. 

A semantic caching strategy is presented in \cite{semanticDataCachingAndReplacement} for client-server database systems. Semantic information  in associative query specifications are used   to organize  the cache at clients that is a composition of multiple semantic regions. A semantic region is a group of semantically related tuples. Semantic regions are defined dynamically when queries are made. Cache replacement is done based on the semantic regions. All  tuples inside a semantic region that is selected to be replaced are discarded from the cache.

All in all, there are existent solutions to decrease latency in mobile networks by caching. Most of them cache content after the first time it is served to a user. They locate caches in different places in the network and they have various algorithms for perceiving duplicate data and storing content in the caches. However, they are not able to serve any content from the caches if the content has not been requested before. Only proactive methods such as P-UPP are capable of serving data from the caches on the first request by prefetching the data before the content is requested. P-UPP predicts user events based on user preference profiles and does not respond to user events actively. 

\section{Semantic Caching and Prefetching Framework}
\label{sec:semanticcaching}

The framework we propose in this paper brings the ability to prefetch data from the Internet each time users make requests. It makes inferences on  requests and predicts what the subsequent requests of the users will be. Depending on these predictions, the inferred contents are prefetched from the Internet and stored in  caches. When the user requests a content that has been cached before thanks to previous inferences, that content is served directly from the caches without retrieving it from the Internet. The framework is composed of two main parts. One part is carried out on user equipment (UE) and the other one is done at cloudlet attached to radio access network (RAN) or Evolving Packet Core (EPC) elements of 4/5G networks.

\subsection{On User Equipment}

The system is based on the Internet Protocol. The meta-data describing a request has to be generated on UE and placed in the hop-by-hop options header of the generated IP datagram. The meta-data  describe, define and/or annotate the data it accompanies. Meta-data are useful if they are created such that it will be used operatively. Our purpose for generating meta-data is using them to predict subsequent requests of users. The meta-data language is determined as Web Ontology Language (OWL) \cite{mcguinness2004owl} due to its expressiveness and inference capability. 

Hop-by-hop options header of the IPv6 is suitable for transportation of meta-data. This header is processed at each hop in the network that is required for in-network caching. This header has a length field allowing variable sized data. In a hop-by-hop options header, there could be many options each having 8 bit length fields. Individual options may have up to 255 bytes of data but upper limit for the Hop-by-Hop Options header size is 2048 bytes which may hold 2030 bytes of data. 

Other than generation of the meta-data and their placement in the IPv6 options headers, no other operation is needed on UE. Packets will travel through network as usual and hop-by-hop header will be used by routers, network elements and the cloudlets for inferring subsequent requests. 

As an open research challenge tools for meta-data generation and annotation are required. We set it outside the scope this paper since we concentrate on the overall framework.

\subsection{At Edge Cloud or Cloudlets}
\label{atTheMobileEdgeCloud}

At the mobile edge cloud or cloudlets attached to network elements, the caching application runs. It processes every request generated by  users and takes necessary prefetching and caching decisions. It keeps a meta-data map for managing the cache.  When IPv6 packets arrive at the  edge cloud, their hop-by-hop options header is retrieved by the caching application. If there are more than one options, their data is appended to each other to form the meta-data. Then, the meta-data are searched in the meta-data map. If they are not found in the cache, this means that the content is already cached and ready to be served to the user. In this case, the request will not be sent to the Internet and the content is served to the user directly from the cache. If the content is not found in the cache, on the other hand, the request will be sent to the Internet. When the resulting content arrives at the mobile edge cloud, the caching application will store the content in the cache and puts its cache address and the meta-data to the meta-data map.

Whether or not the requested content found in the cache, caching application will predict subsequent requests of the user by running a semantic inference algorithm and will store them in the cache after prefetching them from the Internet. This part can be started right after the arrival of the meta-data and is independent from serving the requested content to the user. Semantic inference algorithm takes the meta-data of the current and past requests as input and generates meta-data for the contents that are foreseen to be requested by the user. Since the meta-data are represented with Web Ontology Language (OWL)  \cite{mcguinness2004owl}, it is easy to find related contents. For example, the director of a movie can easily be found using existing ontologies and meta-data of the movie. While an inference algorithm may choose the director of the movie some other may choose actors and actresses of the movie to be prefetched. We do not specify a semantic inference algorithm because it is not in the scope of this paper. Designing better inference algorithms for increasing hit ratio in this framework is an open research challenge.  In \cite{semanticInferenceEMarket}, an example semantic inference algorithm that is used in the prediction of next e-marketplace activities can be found.

After the employed semantic inference algorithm infers the subsequent request of user(s), caching application prefetches those contents from the Internet to the cache. If the inferences turn out to be on target and  users will request prefetched contents that  will directly be served from the cache. Consequently, the request will be satisfied faster. Designing semantic inference algorithms is another research challenge that we set outside the scope of this paper since it is a well-studied topic in other domains. However, it is crucial for the proposed framework to employ a solid inference algorithms to increase hit ratio. In Section \ref{sec:validationAndResults}, we present the ontology and the semantic inference algorithm we used for validating the framework.


\subsection{Location of The Caches}
\label{sectionLocationOfTheCaches}

This framework aims to be ready for subsequent requests of the users by predicting them through inference. One of the  purposes of this framework is serving a cached content to many users after its first usage. However, the main and more important purpose is to serve a cached content to a specific user at the first time it is demanded. Therefore, deployed caches do not need to be located at a place that is at the core of the mobile network. EPC loses the advantage of being at the center of the network and controlling too much traffic. On the other hand, eNodeBs serve to smaller number of users compared to the core network which becomes an advantage in our case.

Prefetched content may never be requested and the bandwidth consumed for prefetching may be wasted.  Locating the cache at  eNodeBs will cause extra bandwidth usage in  backhaul links. However, if the cache is located at  EPC, then  links between RAN and EPC will not be used for prefetching. If the backhaul links are under-utilized,  locating caches at eNodeBs will be better. Assuming  all the other conditions are equal, placing caches  in the RAN will absolutely decrease, compared to caching within EPC, the retrieval time of the content because a link is removed from the path of the data. Therefore, placing caches at RAN has a great advantage when the main purpose is decreasing the end-to-end latency for subsequent requests.

The above reasons indicate that placing caches at RAN might be a better choice for mobile networks when the backhaul links between RAN and EPC are not used with its full potential. However, if the backhaul links are already congested, placing caches at  EPC should be preferred.  A hybrid system can also be designed. Depending on the available budget, caches can be located at both options and a collaborative operation may be enabled. Designing an adaptive hybrid caching system is an open research challenge.

\subsection{Collaboration Opportunities Facilitated by the Framework}

Beyond the meaning of the user requests and responses, the framework facilitates incorporating any meta-data related to  user, context or equipment.  For instance, the capability of the device, the social profile of the user, or the preferences of the user in any domain can be embedded into the options headers.  Collaborative caching can be  enabled in this fashion. Several use-cases can be as follows:
\begin{itemize}
\item Users have social networks that can be represented using the friend-of-a-friend ontology (FOAF). They also have domain specific preferences, for instance movie preferences. Collaborative filtering can be employed when inferring subsequent user request about movies. When friends in  a cell make requests, the preferences can be correlated and exploited in caches collaboratively. 
\item Assume  a family willing to watch a movie together. By employing geotags and movie preferences, not only collaborative caching can be employed but also the bandwidth of all family members can be aggregated for faster streaming.
\item Device capabilities may be used for transcoding multimedia before caching. If the device does not support high-definition videos, down-coded videos can be prefetched and cached.  Full-rate movies stored in caches can also be transcoded after a request comes by utilizing edge computing.
\item When mobility patterns of multiple users can be predicted and group mobility is inferred by considering the geotags and their alterations in time, collaborative caching strategies can be employed. For instance, if a user will be handed over to another cell that employs a distinct cache, the prefetched contents can be redirected to the subsequent cell's cache.
\item Collaborative filtering can be used to identify viral videos and content by exploiting the preference profiles of users embedded in the extension headers. 
\end{itemize}

\section{Validation and Results}
\label{sec:validationAndResults}

A mobile network is emulated using Mininet.  In mobile networks, mobile device users may request any kind of content at any time. But making semantic inferences on an unlimited domain requires use of many ontologies and great amount of work on the semantic inference algorithm. Since finding the semantic inference algorithm is not in the scope of this paper, we have chosen a single ontology and determined a simple algorithm that is useful in a limited domain. The ontology we have used is DBPedia Ontology. We have limited the domain to Wikipedia pages of people and television series. When mobile device users request a Wikipedia page of a person, the semantic inference algorithm chooses Wikipedia page of the spouse of the person as the content to be cached. Similarly, if the user requests the Wikipedia page of a television series, the semantic inference algorithm chooses the Wikipedia pages of the stars of the television series as the contents to be cached. To make the test cases reflect the real life, we have collected the search requests from (around 20) students of WINS Lab at METU. Users make 20 to 30 requests from Wikipedia. We compare the proposed framework with the legacy caching approach called as Traditional Caching where no prefetching or inference is employed. The framework is named as Semantic Caching in the sequel.

\subsection{Impact of the Number of Users}
\label{differentUserCount}

\begin{figure*}
    \centering
     \begin{subfigure}{0.48\textwidth}
		\includegraphics[scale=0.6]{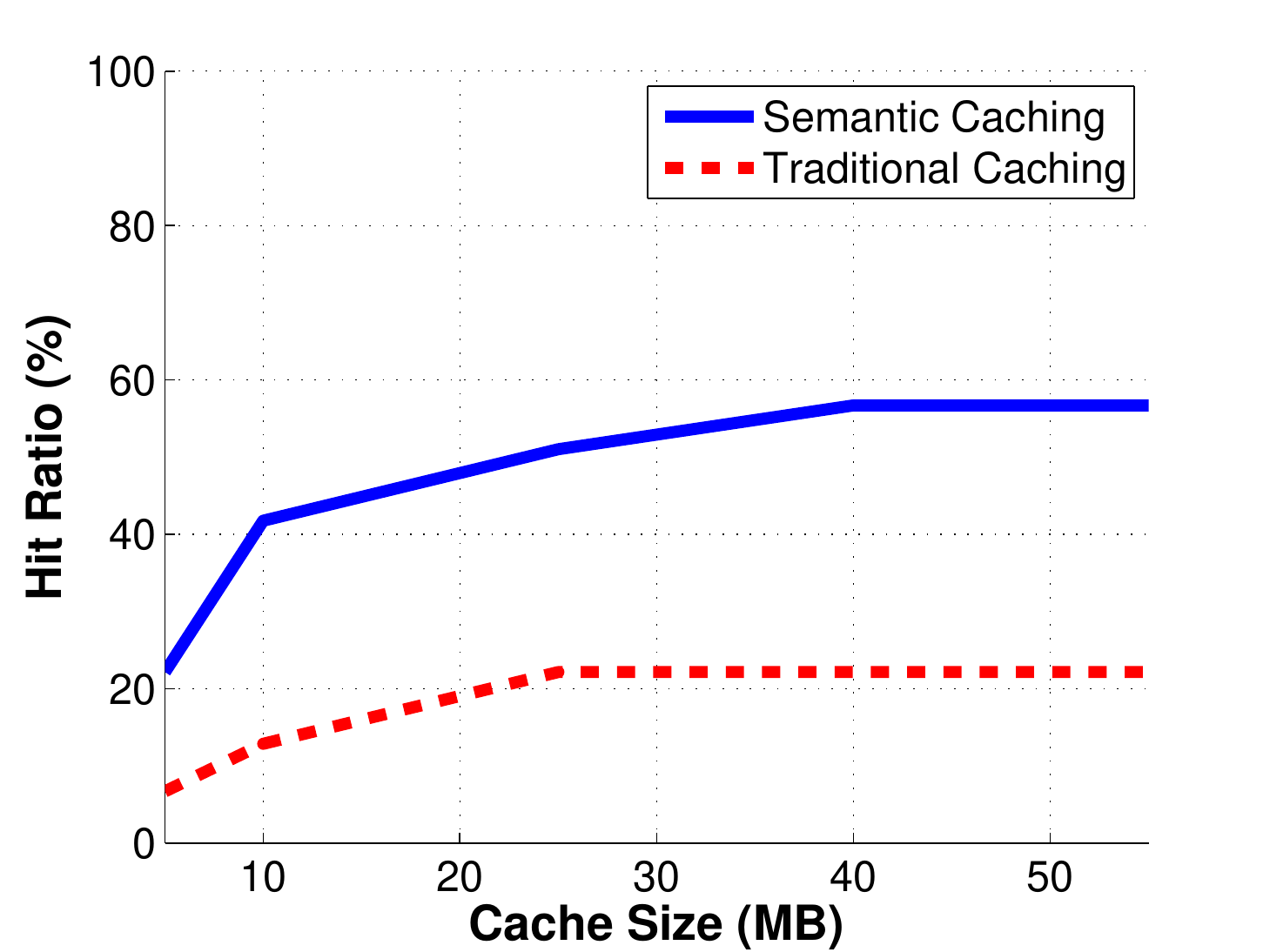}
	\caption{Hit ratio}
	\label{effectsUserCountHitRatio}
    \end{subfigure}   
    \begin{subfigure}{0.48\textwidth}
	\includegraphics[scale=0.6]{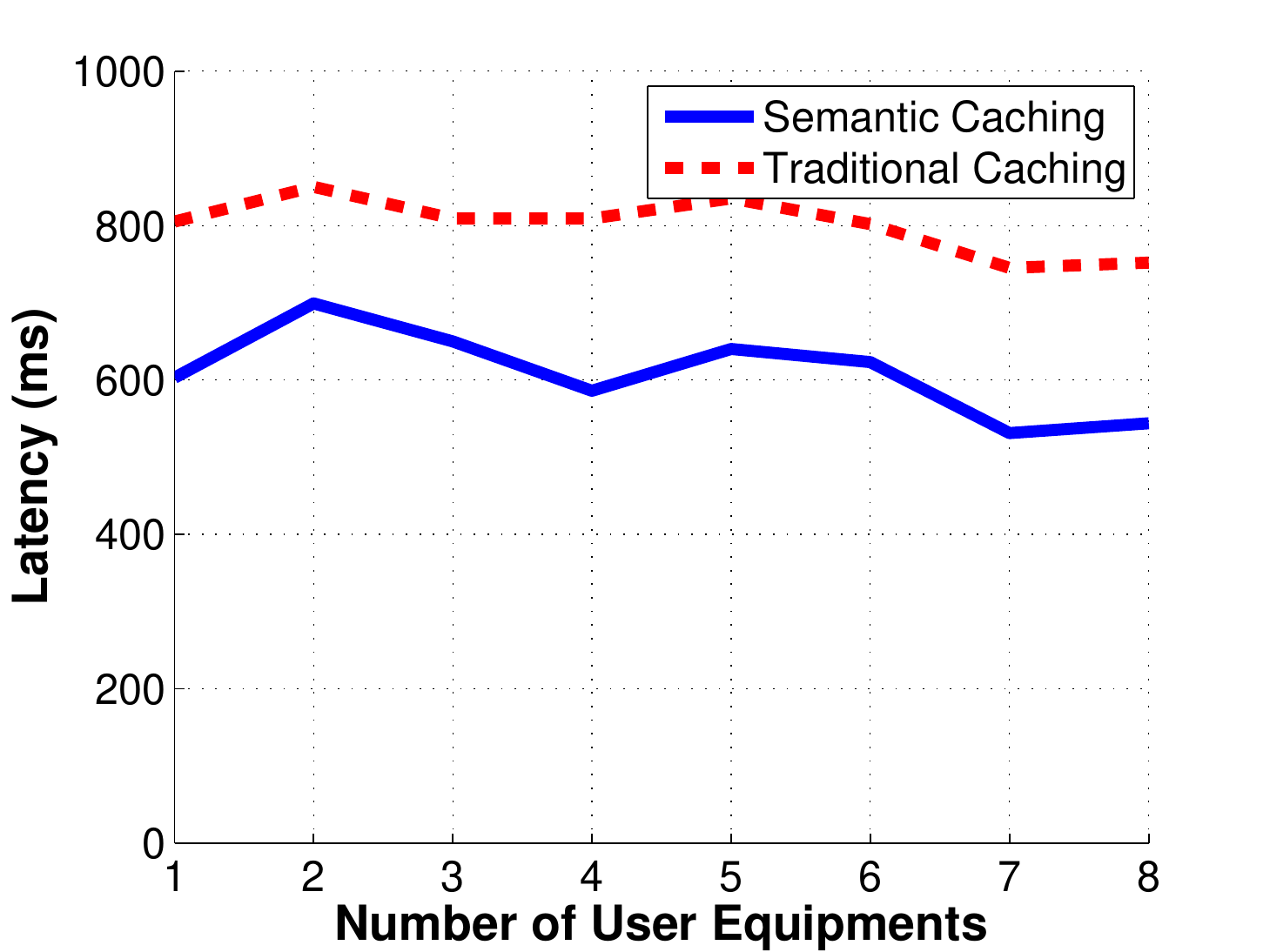}
	\caption{Latency}
	\label{effectsUserCountLatency}
    \end{subfigure}%
    \caption{Impact of the number of users requesting content in a specific domain on  hit ratio and  latency.}
	\label{effectsUserCount}
\end{figure*}

The impact of the number of users on hit ratio and latency are shown in Figure \ref{effectsUserCount}. The cache is located at  the eNodeB. We considered a small number of users in this test case since we assume that there will be a limited number of users generating requests in a cell within the selected domain. Semantic Caching provided more than 50\% hit ratio where as traditional Caching provided  22\%. Although the hit ratio for Semantic Caching has shown just a slight increase as the number of users increase, the hit ratio for Traditional Caching demonstrated hit ratios as low as 7\% with small number of users. Semantic Caching provided 615\% increase in the hit ratio with one user. The minimum increase in the hit ratio that Semantic Caching provided in this group of test cases is 148\% which is a substantial increase. The decrease in the latency when Semantic Caching is preferred over Traditional Caching fluctuates around 24\% and never goes under 17\% as can be seen in Figure \ref{effectsUserCountLatency}.

\subsection{Impact of Cache Size}
\label{differentCacheSize}

\begin{figure*}
    \centering
    \begin{subfigure}{0.48\textwidth}
	\includegraphics[scale=0.6]{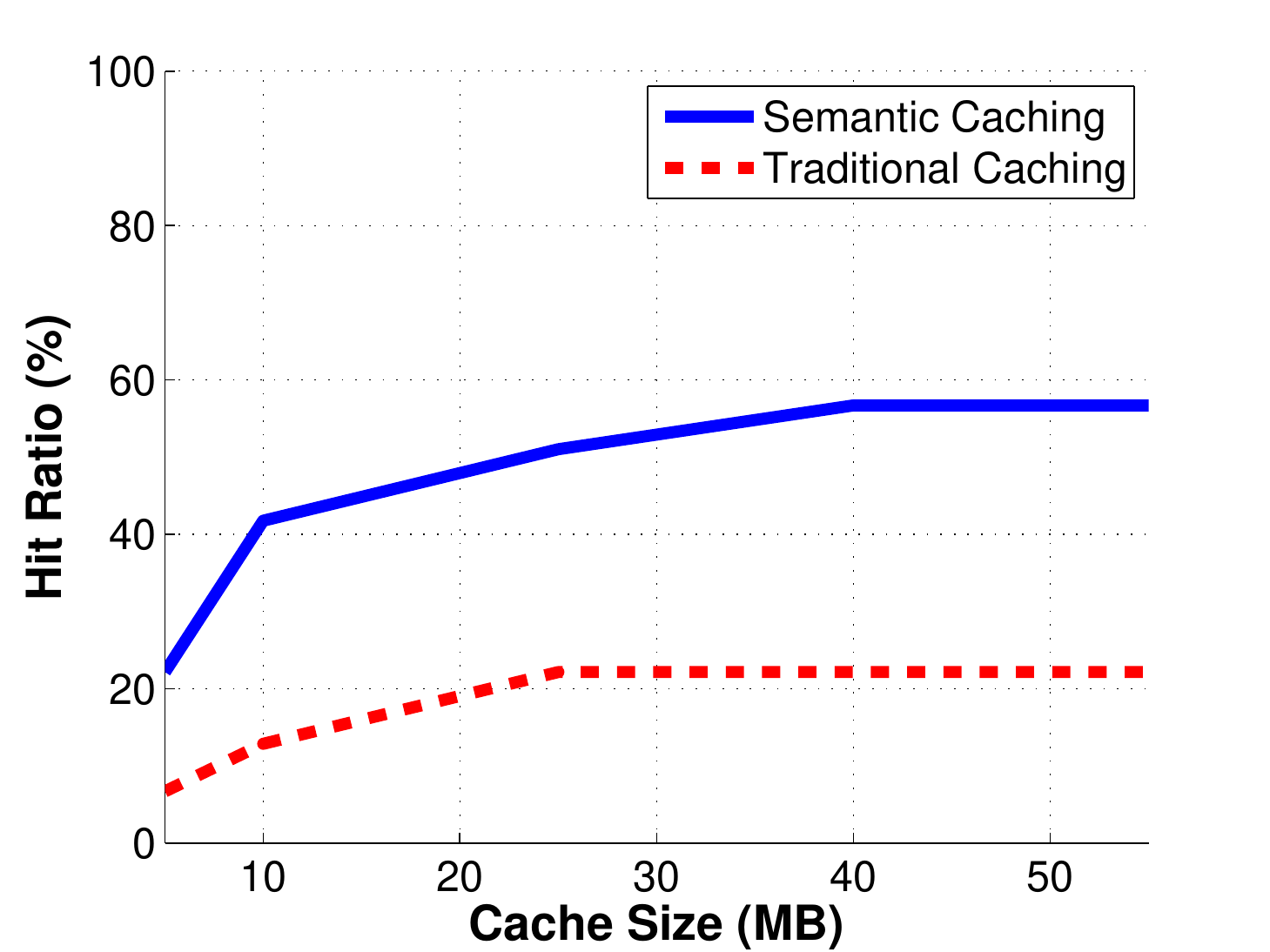}
	\newline
	\caption{Hit ratio}
	\label{effectsCacheSizeHitRatio}
    \end{subfigure}%
    \begin{subfigure}{0.48\textwidth}
	\includegraphics[scale=0.6]{1.pdf}
	\newline
	\caption{Latency}
	\label{effectsCacheSizeLatency}
    \end{subfigure}%
    \caption{Impact of cache size on hit ratio and latency.}
	\label{effectsCacheSize}
\end{figure*}

Figure \ref{effectsCacheSize} shows the hit ratio and latency results as the cache size changes. The results have shown that the hit ratio increases as the cache becomes larger and this is valid for both Semantic Caching and Traditional Caching as can be seen in Figure \ref{effectsCacheSizeHitRatio}. However, Semantic Caching exhibited up to 56\% hit ratio whereas Traditional Caching has never provided a hit ratio more than 22\%. The hit ratio for Semantic Caching was always greater than the hit ratio for Traditional Caching. When the cache is smaller, the increase in the hit ratio was greater as the cache is enlarged. After a certain size, enlarging the cache did not improve the hit ratio.  In Figure \ref{effectsCacheSizeHitRatio}, it can be seen that Semantic Caching has provided between 130\% and 230\% increase in the hit ratio. This amount has taken its maximum value at the smallest cache size, which is 5 megabytes since Traditional Caching has provided a hit ratio as low as 6\% while Semantic Caching exhibited hit ratio more than 20\%. Since the hit ratio for both Semantic Caching and Traditional Caching has increased as the cache size becomes larger, the latency of both methods has decreased with larger caches as expected  (Figure \ref{effectsCacheSizeLatency}). 

\subsection{Impact of Cache Location}
\label{diferentCacheLocation}
\begin{figure*}
\begin{subfigure}{.33\linewidth}
\centering
\includegraphics[scale=0.35]{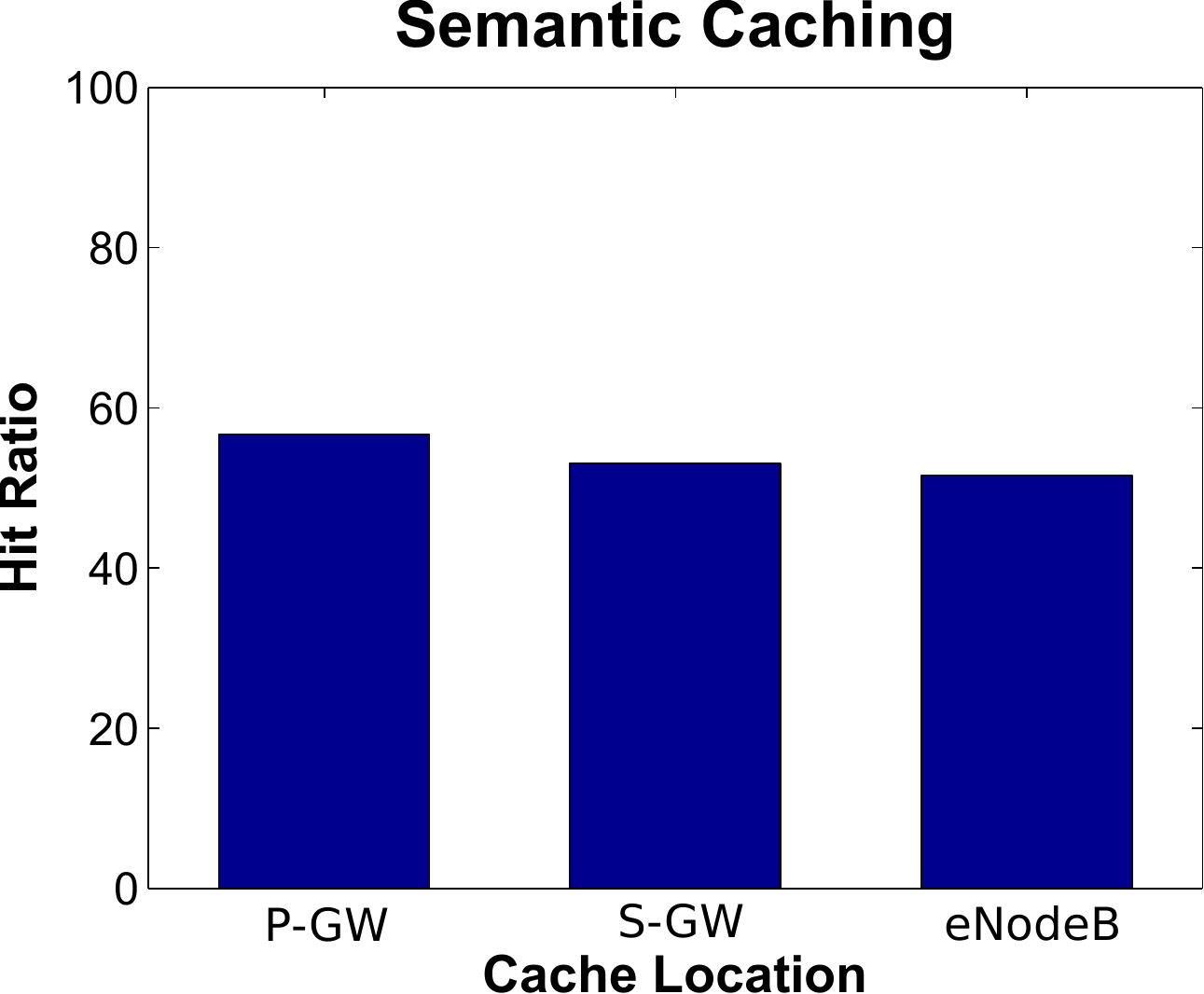}
	\caption{Hit ratios for Semantic Caching.}
	\label{effectsCacheLocationHitRatioA}
\end{subfigure}%
\begin{subfigure}{.33\linewidth}
\centering
\includegraphics[scale=0.35]{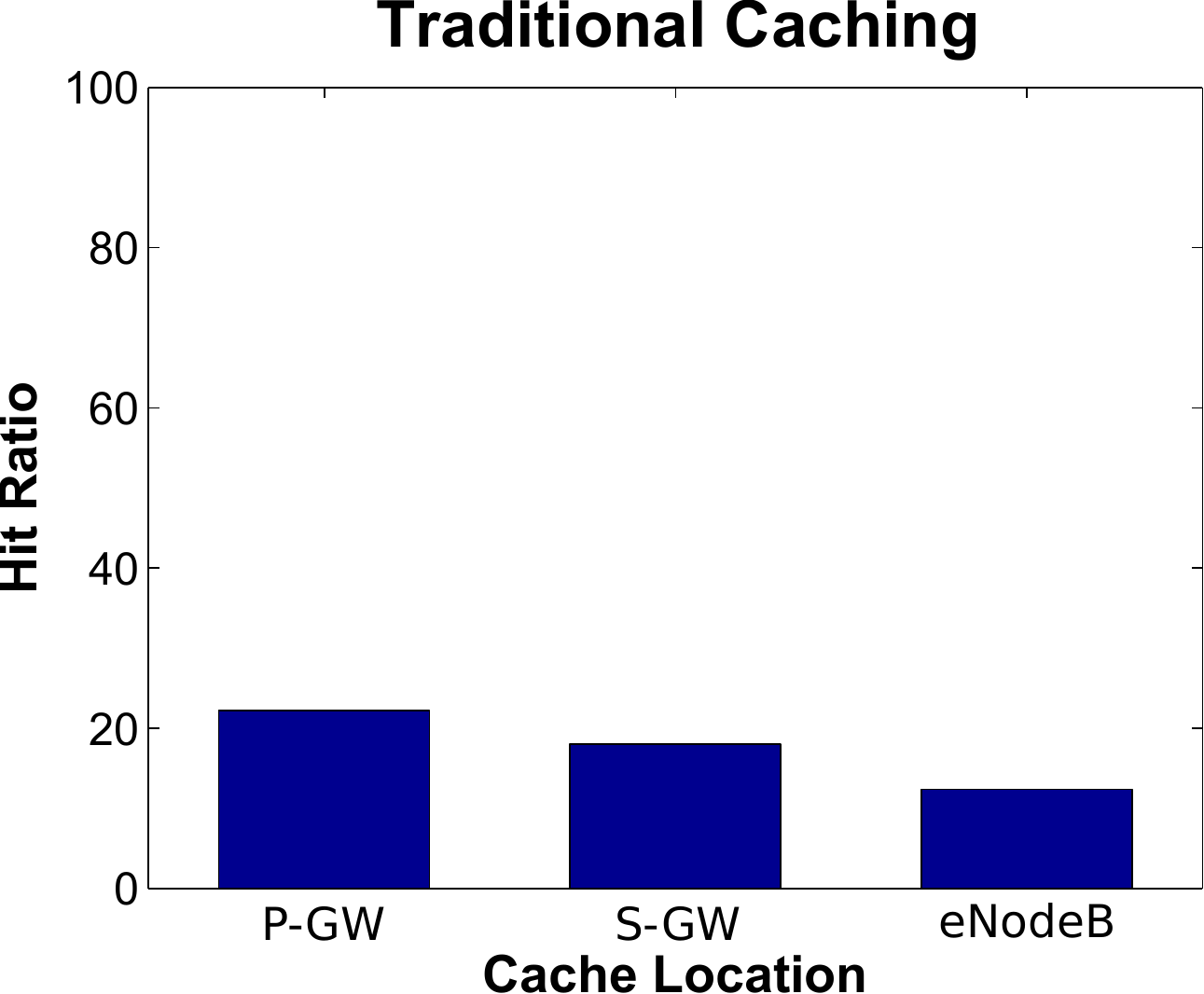}
	\caption{Hit ratios for Traditional Caching.}
	\label{effectsCacheLocationHitRatioB}
\end{subfigure}
\begin{subfigure}{0.33\linewidth}
\centering
\includegraphics[scale=0.35]{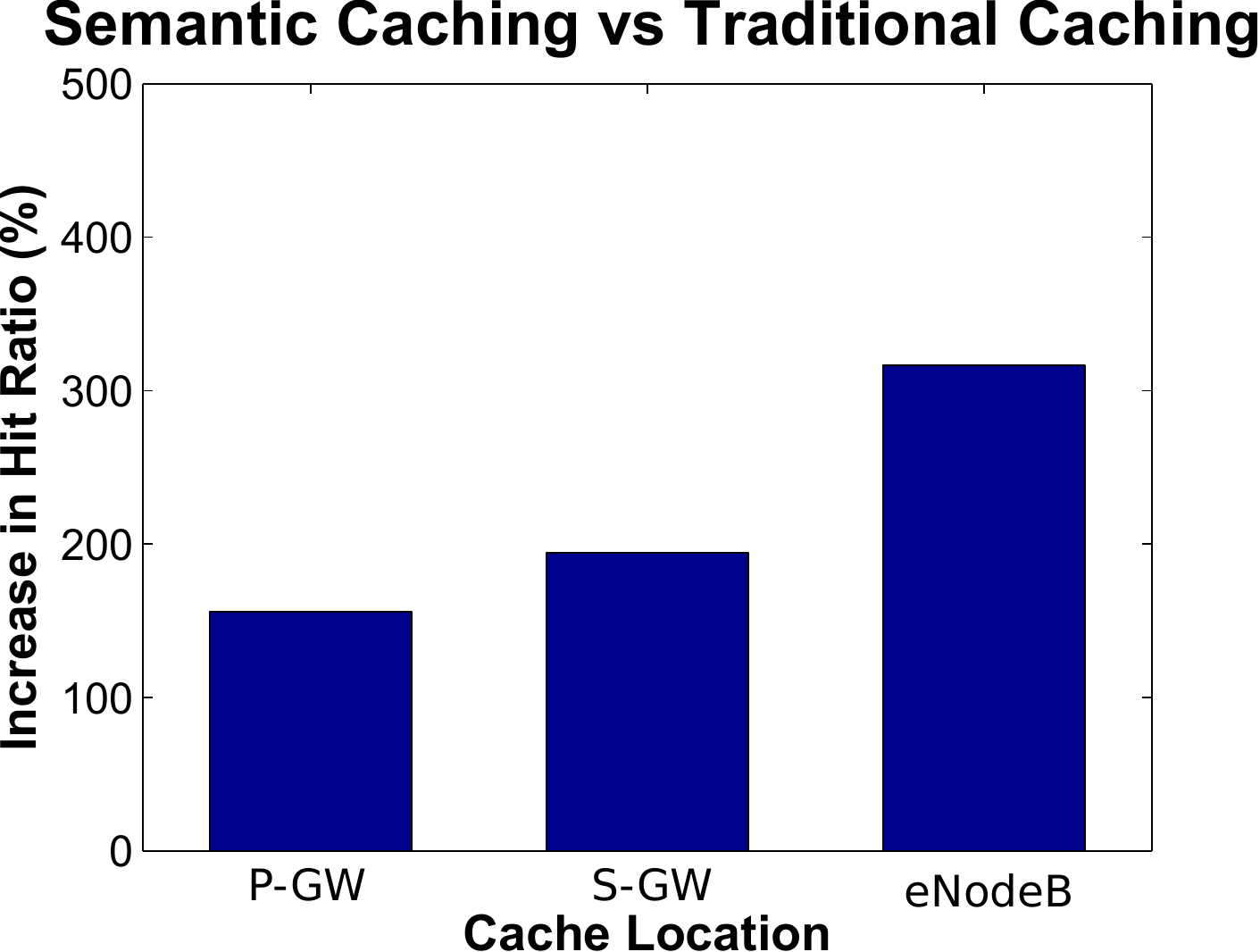}
	\caption{Increase in hit ratio (\%).}
	\label{effectsCacheLocationHitRatioC}
\label{fig:sub3}
\end{subfigure}\\[1em]
\begin{subfigure}{.33\linewidth}
\centering
\includegraphics[scale=0.35]{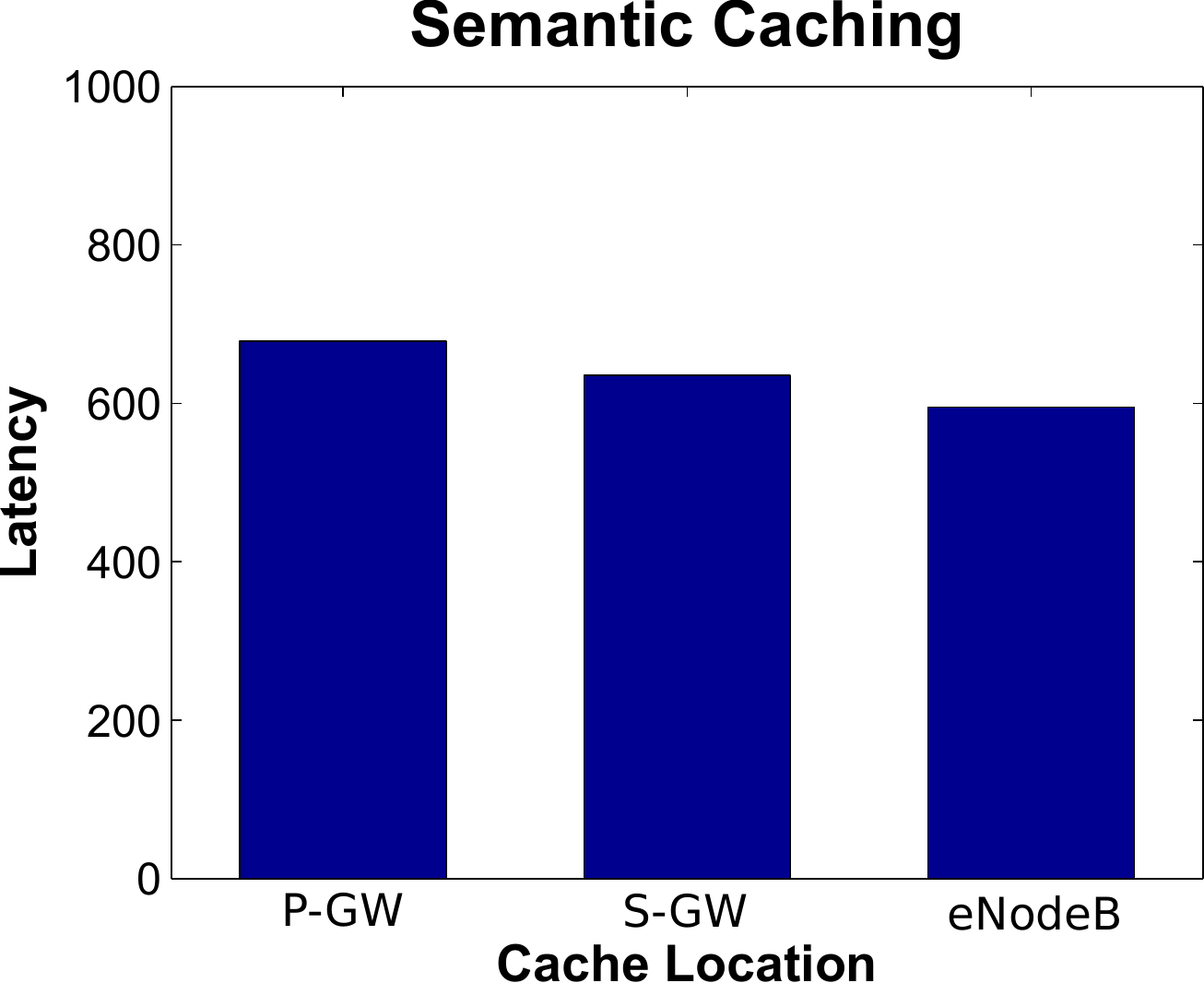}
	\caption{Latency  for Semantic Caching.}
	\label{effectsCacheLocationLatencyA}
\end{subfigure}%
\begin{subfigure}{.33\linewidth}
\centering
\includegraphics[scale=0.35]{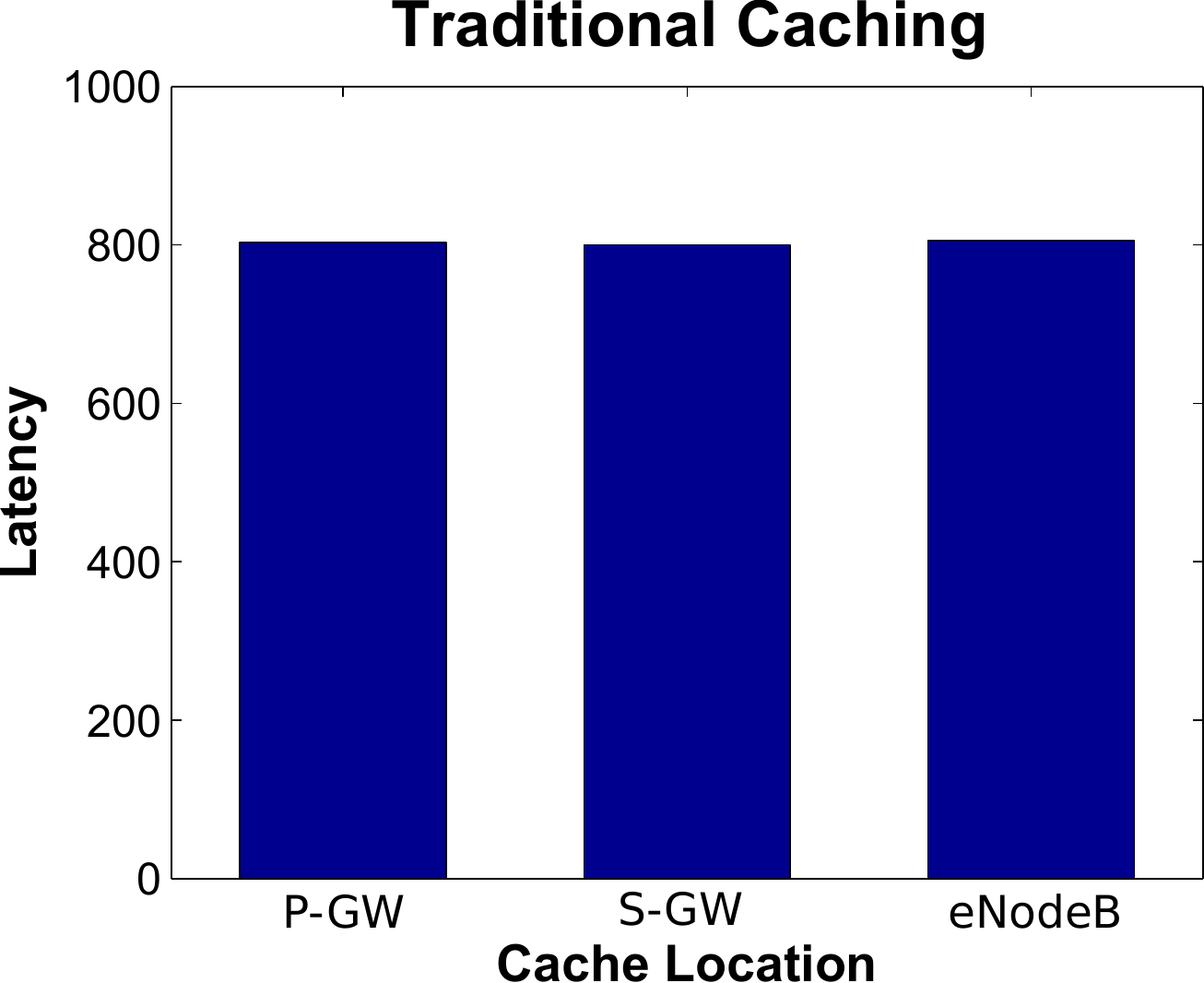}
	\caption{Latency  for Traditional Caching.}
	\label{effectsCacheLocationLatencyB}
\end{subfigure}
\begin{subfigure}{0.33\linewidth}
\centering
\includegraphics[scale=0.35]{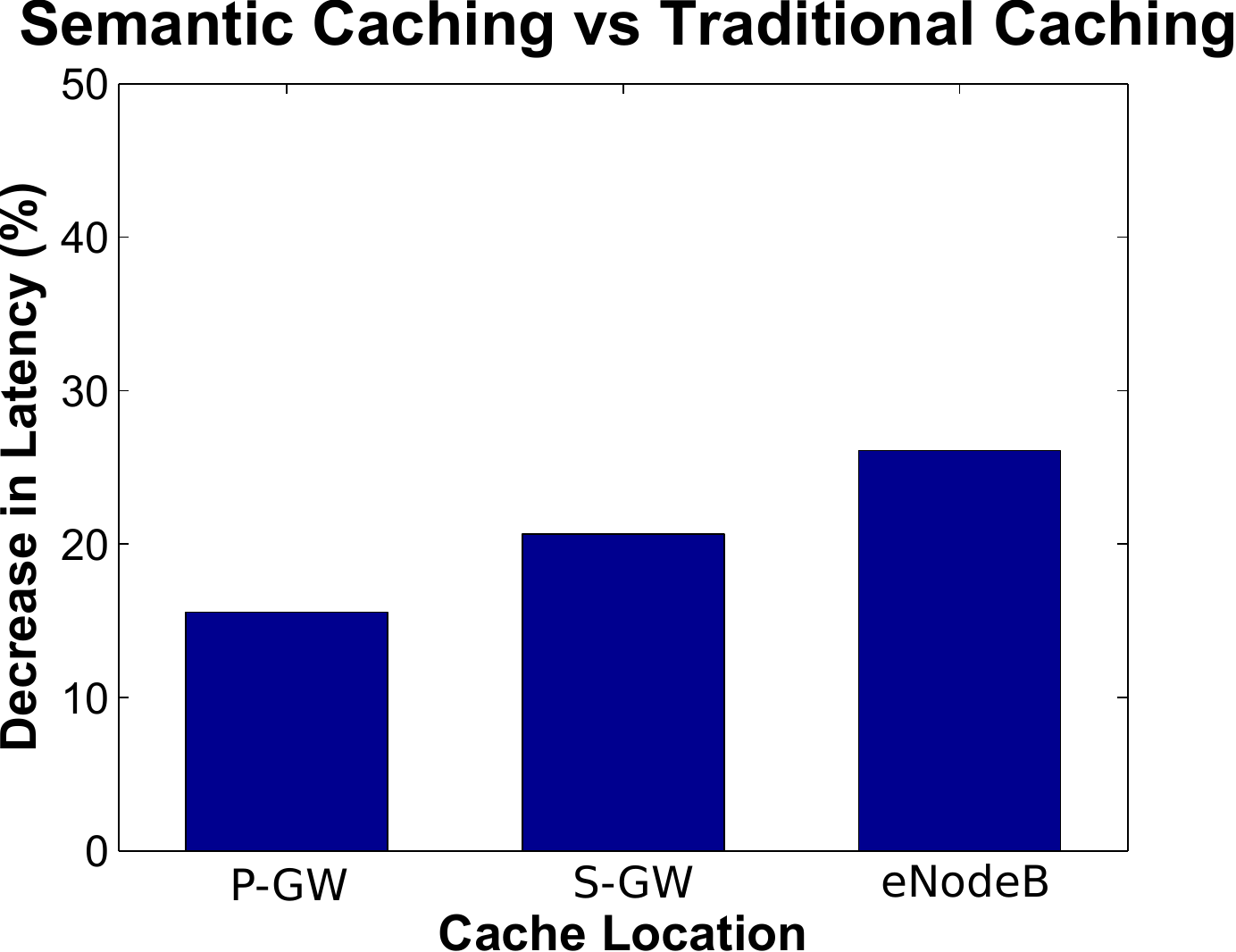}
	\caption{Decrease in latency (\%).}
	\label{effectsCacheLocationLatencyC}
\end{subfigure}
    \caption{Impact of cache location on hit ratio and latency (ms).}
	\label{effectsCacheLocation}
\label{fig:test}
\end{figure*}

In Figure \ref{effectsCacheLocation}, we have compared the hit ratios for Semantic Caching and Traditional Caching with different positions for cache deployment. Semantic Caching and Traditional Caching have shown similar reactions to changing cache location. Both of them provided highest hit ratio when the cache is located near P-GW. The lowest hit ratio is provided when the caches are located near eNodeB as expected. The reason is that when the cache is located in a central place, the expected hit ratio is higher. 

Semantic Caching has provided up to 56\% hit ratio whereas Traditional Caching did not provide any hit ratios higher than 26\% as seen in Figures \ref{effectsCacheLocationHitRatioA} and \ref{effectsCacheLocationHitRatioB}, respectively. Using the data in these two figures, we have calculated the increase in the hit ratio that will be achieved when Semantic Caching is preferred over Traditional Caching and depicted in Figure \ref{effectsCacheLocationHitRatioC}. Here, we can see that the improvement in hit ratio of Semantic Caching increases as the cache gets closer to UEs. The reason is that Semantic Caching aims to serve cached content to a specific user, which will be an advantage when the caches are not located in a central place, as explained in Section \ref{sectionLocationOfTheCaches}. 

In Figure \ref{effectsCacheLocation},  we present the comparison of latency results with different cache deployment locations. Semantic Caching provides lower latency values when the caches are located closer to UEs. This is an expected result when we consider that the hit ratio increased as the caches get closer to UEs as explained above. Similar to the hit ratio results, latency values for Semantic Caching is always lower compared to Traditional Caching as seen in Figures \ref{effectsCacheLocationLatencyA} and \ref{effectsCacheLocationLatencyB}. The decrease in the latency that is achieved when Semantic Caching is preferred rather than Traditional Caching is depicted in Figure \ref{effectsCacheLocationLatencyC}. This figure shows that the improvement in the latency that Semantic Caching provides increases as the cache gets closer to UEs.

\subsection{Cost}
The above mentioned benefits are attained at a cost. Actually, there are two drawbacks that Semantic Caching brings. The first one is that meta-data describing the user request are added to the IPv6 packet inside a hop-by-hop options header and travel through the mobile network. In our test cases, 1.24 kilobytes of data is transmitted through the network for each user which means 0.01\% increase in the amount of transmitted data. The second and the main drawback of Semantic Caching is that additional data traffic is generated when prefetching contents because some of those contents will never be requested by  users. In our test cases, we have observed 39\% useless data traffic that are prefetched but not requested. The percentage of the redundant data traffic depends on the semantic inference algorithm and designing better semantic inference algorithms  -that we set outside the scope of this paper albeit it is a significant research challenge- will decrease the generated redundant data traffic.

\subsection{Overall Performance}
\begin{figure*}
    \centering
    \begin{subfigure}{0.5\textwidth}    \centering
	\includegraphics[scale=0.6]{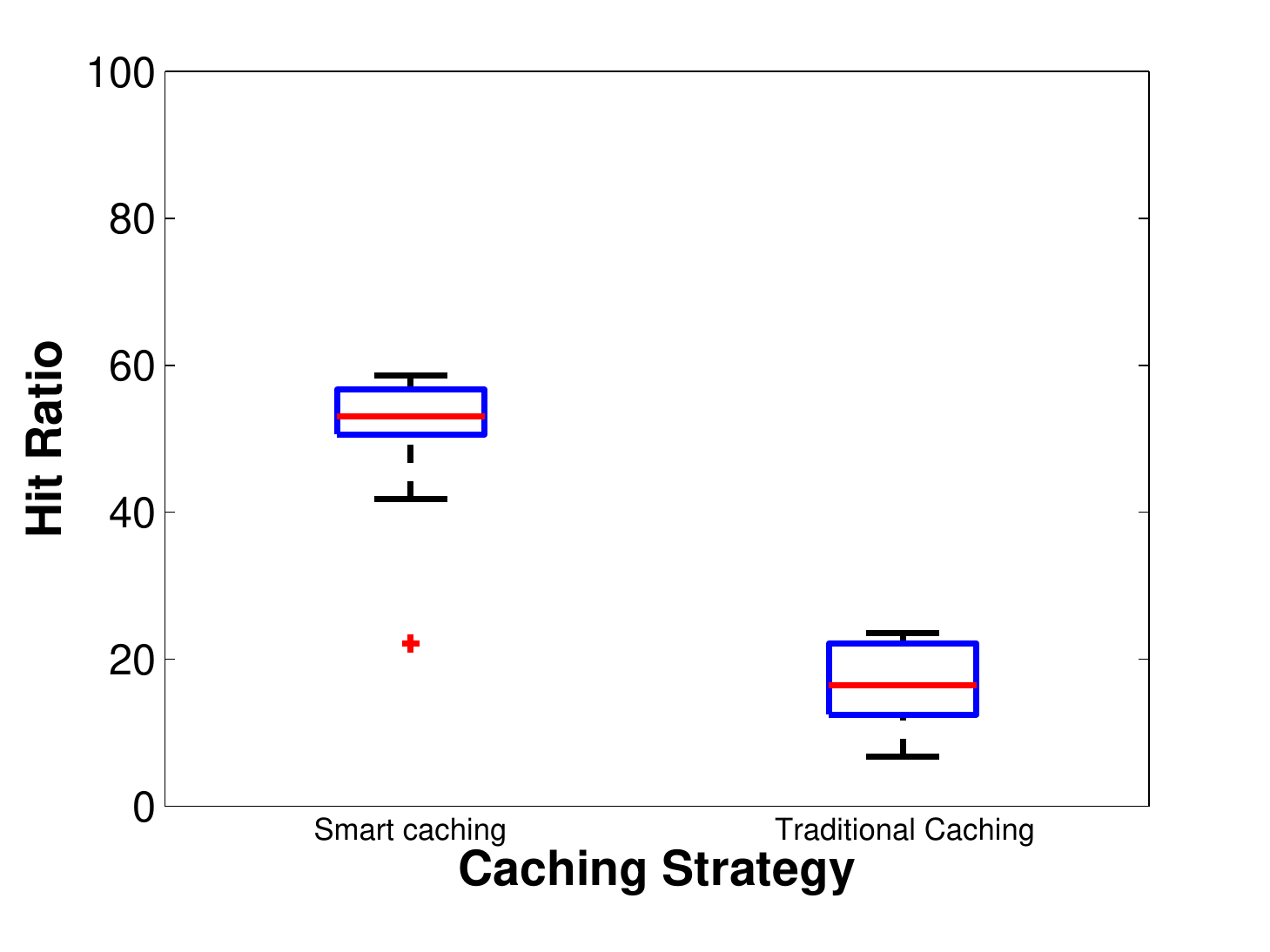}
	\caption{Hit ratio}
	\label{overallPerformanceBoxA}
    \end{subfigure}%
    ~ 
    \begin{subfigure}{0.5\textwidth}    \centering
	\includegraphics[scale=0.6]{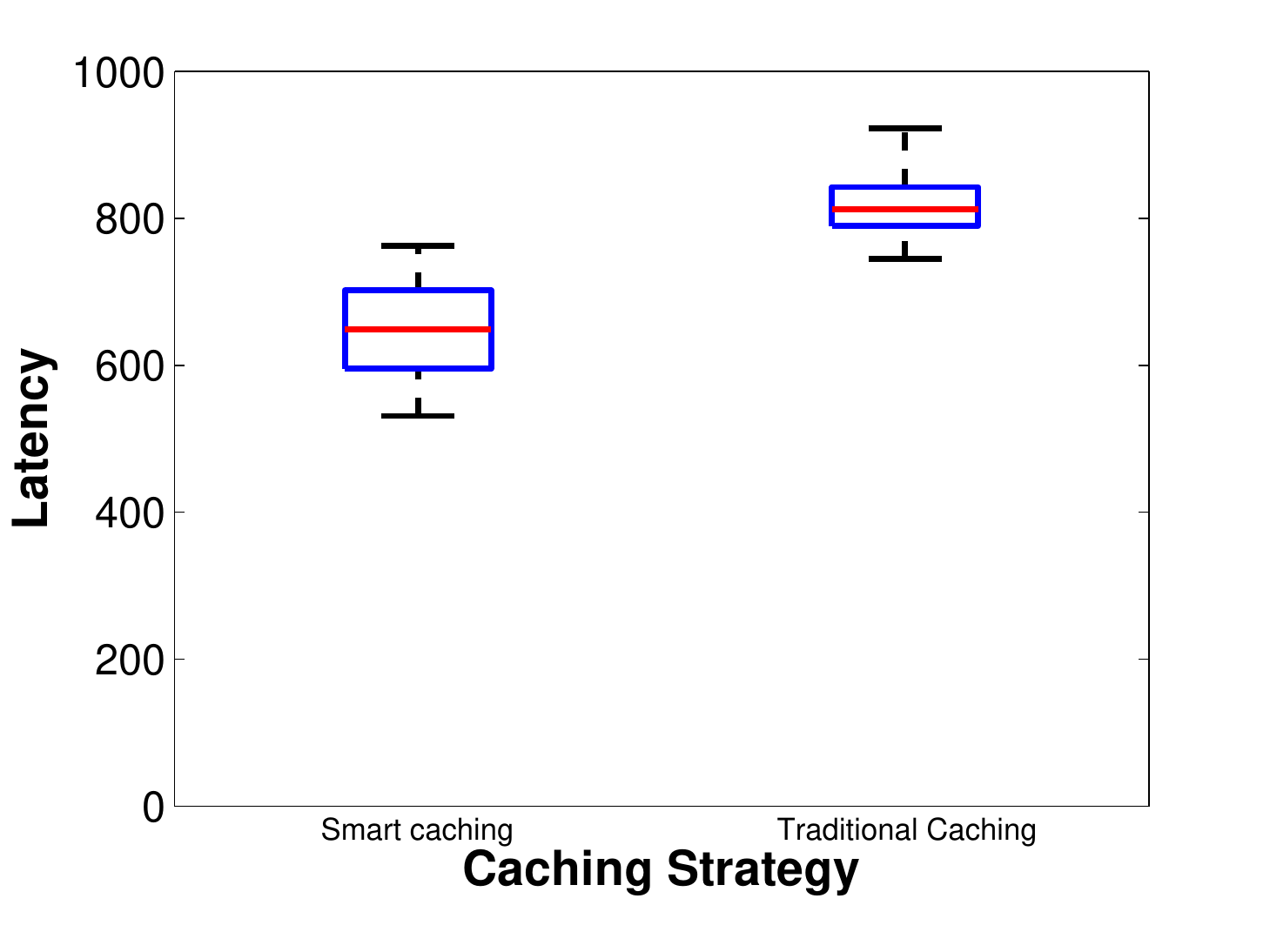}
	\caption{Latency (ms)}
	\label{overallPerformanceBoxB}
    \end{subfigure}
    \caption{Overall performance comparison of Semantic Caching and Traditional Caching.}
	\label{overallPerformanceBox}
\end{figure*}

To measure the overall performance of the framework, we have taken the average of the results obtained in all of the 31 test cases. On the average, Semantic Caching exhibited 52\% hit ratio whereas Traditional Caching has provided only 15\% hit ratio as seen in Figure \ref{overallPerformanceBoxA}. Similarly, Semantic Caching exhibited improvement in  latency compared to Traditional Caching that can be seen in Figure \ref{overallPerformanceBoxB}. In fact, Semantic Caching provided 20\% lower latency compared to Traditional Caching on the average. To be able to provide such improvements in hit ratio and latency, Semantic Caching caused 39\% useless data traffic between the network element where the cache is located and the Internet.

\section{Conclusion}

In this paper, we propose a proactive in-network semantic caching framework for 5G mobile networks that is able to predict future user requests and store them in the caches after prefetching them. It  serves contents from the caches even at the first time they are demanded by employing semantic inference technologies. Its main difference from existing caching strategies is that on each user request it makes inferences on the  request(s) and prefetches contents that are predicted to be the subsequent requests of the user. The aim of proposing this strategy is helping to reduce the latency by serving more contents from the caches.

In all of the 31 test cases we have created, Semantic Caching provided higher hit ratios and lower latency compared to Traditional Caching. In fact, Semantic Caching exhibited 52\% hit ratio whereas Traditional Caching provided 15\% hit ratio, on the average. Moreover, Semantic Caching provided 20\% lower latency compared to Traditional Caching, on the average. The decrease in the latency is higher when the caches are located near the eNodeB and lower when the caches are located near the P-GW. In addition to these improvements, Semantic Caching brings some drawbacks. The main drawback it brings is that while prefetching contents additional data traffic is generated. In fact, 39\% of the generated data traffic was used for contents that are never served to the users. As  caches get closer to  user equipments, more links are used while prefetching. When the caches are located near P-GW, only the links between P-GW and the Internet are used. However, when the caches are located near eNodeB, all the links between eNodeB and the Internet are used. If the links between eNodeB and S-GW are generally congested in a mobile network, it is not a good idea to place  caches in eNodeBs. If all of the links in the mobile network, on the other hand, is not congested most of the time, then placing the caches near  eNodeBs can be preferred   to minimize  latency.


%


\section*{Acknowledgment}
This work is partially supported by TUBITAK under the grant number 115C064.


\ifCLASSOPTIONcaptionsoff
  \newpage
\fi

\end{document}